# Fermi level position, Coulomb gap, and Dresselhaus splitting in (Ga,Mn)As

S. Souma[1*], L. Chen[1], R. Oszwałdowski[2], T. Sato[3], F. Matsukura[1,4,5], T. Dietl[1,6,7], H. Ohno[1,4,5], & T. Takahashi[1,3].

[1] *WPI Research Center, Advanced Institute for Materials Research, Tohoku University, 2-1-1 Katahira, Aoba-ku, Sendai 980-8577, Japan*

[2] *Department of Physics, South Dakota School of Mines and Technology, Rapid City, SD 57701, USA*

[3] *Department of Physics, Tohoku University, 6-3 Aramaki Aza-Aoba, Aoba-ku, Sendai 980-8578, Japan*

[4] *Center for Spintronics Integrated Systems, Tohoku University, 2-1-1 Katahira, Aoba-ku, Sendai 980-8577, Japan*

[5] *Laboratory for Nanoelectronics and Spintronics, Research Institute of Electrical Communication, Tohoku University, 2-1-1 Katahira, Aoba-ku, Sendai 980-8577, Japan*

[6] *Institute of Physics, Polish Academy of Sciences, aleja Lotników 32/46, PL-02-668 Warszawa, Poland*

[7] *Institute of Theoretical Physics, Faculty of Physics, University of Warsaw, ulica Pasteura 5, PL-02-093 Warszawa, Poland*

Correspondence and requests for materials should be addressed to S. S. (e-mail: s.souma@arpes.phys.tohoku.ac.jp).

Carrier-induced nature of ferromagnetism in a ferromagnetic semiconductor, (Ga,Mn)As, offers a great opportunity to observe novel spin-related phenomena as well as to demonstrate new functionalities of spintronic devices. Here, we report on low-temperature angle-resolved photoemission studies of the valence band in this model compound. By a direct determination of the distance of the split-off band to the Fermi energy $E_F$ we conclude that $E_F$ is located within the heavy/light hole band. However, the bands are strongly perturbed by disorder and disorder-induced carrier correlations that lead to the Coulomb gap at $E_F$, which we resolve experimentally in a series of samples, and show that its depth and width enlarge when the Curie temperature decreases. Furthermore, we have detected surprising linear magnetic dichroism in photoemission spectra of the split-off band. By a quantitative theoretical analysis we demonstrate that it arises from the Dresselhaus-type spin-orbit term in zinc-blende crystals. The spectroscopic access to the magnitude of such asymmetric part of spin-orbit coupling is worthwhile, as they account for spin-orbit torque in spintronic devices of ferromagnets without inversion symmetry.

Since the elaboration of the way of its preparation[1], manganese-doped gallium arsenide, (Ga,Mn)As, has been the most intensively studied dilute magnetic semiconductors for two interrelated reasons[2,3]. First, Mn ions substituted for Ga act as an acceptor and provide holes, which mediate the ferromagnetic interaction among localized Mn moments. The presence of randomly distributed acceptors leads to a substantial disorder and to metal-to-insulator transition (MIT) in the range of hole densities relevant to ferromagnetism. Interplay of carrier-mediated exchange interactions and quantum Anderson-Mott localization results in striking properties, often difficult to describe quantitatively either analytically or by the state-of-the-art *ab initio* methods. This has opened a room for extensive qualitative debates on the character of states in the vicinity of the Fermi energy and on the mechanism accounting for ferromagnetism in this system[4,5].

Second, because of the novel carrier-induced nature of ferromagnetism as well as the compatibility with well-established GaAs-based devices, (Ga,Mn)As and related compounds have become a platform on which a number of new concepts of spintronic devices has been demonstrated[2,3]. In one kind of these devices carrier-concentration dependent ferromagnetism (*e.g.*, spin-FET[6] or spin p-i-n diode[7]) or the presence of spin current (*e.g.*, spin-LED[8,9] and related spin-injection devices[10], domain-wall displacement[11]) was exploited. In another type of functional structures the direction of the magnetization vector was altered[12-15] *via* spin-orbit coupling originating from the electric field brought about either by intra-atomic potentials[16-19] [and giving rise to splitting between heavy/light hole ($\Gamma_8$) and split-off ($\Gamma_7$) valence-band subbands in zinc-blende semiconductors] or by inversion asymmetry leading to the presence of Dresselhaus terms in the band dispersion $E(\boldsymbol{k})$[15,18,20]. Tunnelling anisotropic magnetoresistance[21] is one more example of devices demonstrated first for (Ga,Mn)As.

Photoemission spectroscopy, and its angle-resolved version that we employ in this work have already provided a number of important information, particularly on hybridization of Mn *d*-levels and valence-band *p*-states, and the associated participation of *d*-states in the wave function of carriers at the Fermi level $E_F$ in (Ga,Mn)As[22-26]. These studies have also brought into light issues associated with surface contamination.

In this work, we at first provide information on samples preparation, protocols implemented to avoid contamination as well as on experimental procedure employed in our studies of (Ga,Mn)As by angle-resolved photoemission spectroscopy (ARPES). The obtained spectra are presented, together with results of tight-binding computations. This comparison allows us to assign particular features to individual valence-band subbands, and to demonstrate that $E_F$ is located within the heavy/light hole band in ferromagnetic (Ga,Mn)As. At the same time, density of states (DOS) is considerably depleted at $E_F$, which we take as new evidence for the presence of the Coulomb gap, driven by disorder-enhanced carrier correlations, as predicted by Altshuler and Aronov[27]. By studying a series of samples we have found that the lower Curie temperature $T_C$, the larger DOS depletion, *i.e.,* $T_C$ decreases with approaching the MIT. It is emphasized that such a Coulomb gap cannot be reproduced by the current *ab initio* method, since it results from quantum interference of carrier-carrier interaction amplitudes at the mesoscopic scale; therefore the experimental output on the Coulomb gap is of particular importance. Another new effect revealed by our studies is magnetic linear dichroism (MLD), particularly well resolved for the spilt-off subband. By direct computations of light absorption for transitions from this band to free-electron states we demonstrate that MLD originates from the Dresselhaus spin-orbit terms, and is present for both s and p light polarization. This is intriguing since the MLD observed in ferromagnets by core-level photoemission[28] is active only for p-polarization[29].

**Results**

**Samples and experimental.** (Ga,Mn)As films with 25-nm thickness were grown by the molecular beam epitaxy (MBE) method[1], and then transferred to the ARPES chamber without being exposed to the air by using a portable ultrahigh vacuum (UHV) chamber, which keeps $10^{-10}$ Torr during the transportation (see Methods). This procedure has proved very effective to obtain reliable ARPES data because ARPES is very sensitive to the condition of sample surface[22].

The properties of (Ga,Mn)As are strongly affected by the growth conditions during MBE[30]. To avoid the formation of MnAs precipitates, it is necessary to grow (Ga,Mn)As below 250°C (ref. 1), while the low-temperature growth likely produces anti-site As defects acting as double-donors[31]. It is also known that Mn atoms in GaAs occupy either the Ga sites or the interstitial positions[32]. Ga-substitutional Mn provides a hole by acting as an acceptor, while the interstitial Mn compensates holes by acting as a double-donor and its magnetic moment couples antiferromagnetically with that of the substitutional Mn[33]. To obtain higher metallicity of films, one needs to grow (Ga,Mn)As with less amount of interstitial Mn atoms, whose formation energy decreases with increasing Mn content[34]. Although the post-annealing of films effectively removes the interstitial Mn atoms, the annealing alternatively causes precipitation of Mn atoms and/or Mn-related compounds at the surface[35]. Because ARPES is very surface-sensitive technique, it is crucial to exclude these extrinsic factors that distort ARPES results.

We have grown 25-nm-thick $Ga_{1-x}Mn_xAs$ layers with nominal Mn compositions $x$ between 0.05 and 0.07 on an $n^+$-GaAs (001) substrate through a 100-nm thick $n^+$-GaAs buffer layer at the growth temperature between 230 and 245°C. Magnetic characterization has been performed after the ARPES measurements, and a typical temperature dependence of in-plane magnetization of (Ga,Mn)As with a relatively high $T_C$ above 100 K is presented in Fig. 1a. We find that the higher growth temperature tends to result in the higher values of $T_C$. We prepared a series of (Ga,Mn)As films with $T_C$ between 35 and 112 K in the as-grown state. A reference n-type GaAs film doped with Si concentration of $10^{19}$ cm$^{-3}$ has been grown on an $n^+$-GaAs (001) substrate in the same MBE chamber.

**Location of the Fermi level.** Figure 1c shows the experimental band structure of $Ga_{0.95}Mn_{0.05}As$ ($T_C$ = 101 K) along the $\bar{\Gamma}\bar{X}$ cut obtained by plotting the ARPES-spectral intensity as a function of binding energy and wave vector. We have measured ARPES spectra with the He-Iα resonance line ($h\nu$ = 21.218 eV) with the main polarization vector along the $\bar{\Gamma}\bar{X}$ cut (Fig. 1b) at temperature $T$ = 30 K. We observe three dispersive hole-like bands (A-C) centred at the $\bar{\Gamma}$ point, all of which are ascribed mostly to the As 4$p$ orbitals in GaAs[36] and also in (Ga,Mn)As[24]. Band A has the top of dispersion at the binding energy of 2.9 eV. Bands B and C tend to converge on approaching the $\bar{\Gamma}$ point, and their spectral intensity is markedly suppressed in the region within 1 eV from $E_F$. In the inset with enhanced colour contrast, one can see a signature of band B nearly touching $E_F$. This feature is more clearly seen in the experimental band structure obtained by the second derivative of the momentum distribution curves (MDCs) presented in Fig. 1d. It is obvious that band B almost reaches $E_F$ at the $\bar{\Gamma}$ point, while band C is hardly traced around $E_F$ because of the very faint feature. We also observe another band (band D) directing toward $E_F$ on approaching the second $\bar{\Gamma}$ point.

We compare in Fig. 1e the experimental band structure with our tight-binding band calculations for GaAs (see Methods) along the ΓKX ($k_z$ = 0, solid curves) and XKΓ ($k_z$ = $2\pi/a_0$ dashed curves) high-symmetry lines. There are good correspondences between the experiment and the calculation. Comparison shows that bands B and C are assigned to the bands at $k_z$ = 0, while bands A and D are at $k_z$ = $2\pi/a_0$. This suggests that the ARPES data reflect the electronic states averaged over a wide $k_z$ region in the bulk

Brillouin zone, so that the electronic states along the high-symmetry lines, *i.e.* at $k_z = 0$ and $2\pi/a_0$, have a dominant contribution to the total "one-dimensional" DOS averaged over $k_z$[36,37]. Therefore, the overall agreement between the experimental and the calculated dispersions as seen in Fig. 1e leads us to conclude that bands B and C are the split-off (SO) and light-hole (LH) bands, respectively.

We have investigated the detailed band structure near $E_F$ at the Γ point. As displayed in Fig. 2a, band B shows a "Λ"-shaped spectral-intensity distribution. To see the dispersive feature more clearly, we have subtracted the featureless background from the raw spectrum (Fig. 2b), and show the result and its intensity plot in Figs. 2c and 2d, respectively. Intriguingly, a clear Fermi-edge structure is recognized in the spectrum when the band approaches $E_F$ (Fig. 2c). It is noted that a similar Fermi-edge-like structure is already seen even in the raw spectrum (Fig. 2b), indicating its intrinsic nature. It is stressed here that the present first observation of the Fermi edge in (Ga,Mn)As is a consequence of the careful surface treatment using the UHV sample transfer chamber as well as the high-energy resolution in the ARPES measurement. As seen in Fig. 2c, the spectral intensity at the Fermi edge systematically increases when band B approaches $E_F$, as highlighted by the MDC at $E_F$ (white curve in Fig. 2d). We also observed a similar spectral characteristics in another (Ga,Mn)As sample with lower $T_C$ of 62 K (Figs. 2e and 2f). As shown in Fig. 2g, band B in both samples with $T_C = 62$ K or 101 K shows a good correspondence to the SO band lying ~0.35 eV below the LH/HH (heavy hole) band at the Γ point in GaAs. Thus, the present ARPES results unambiguously demonstrate that the Fermi level resides deeply inside the As-4$p$ valence band. This conclusion is in accord with recent on-Mn resonance photoemission studies[23,24] which confirm that the main spectral weight of Mn $d$ states is at 3.5 eV below $E_F$. Accordingly, taking also into account a large difference in the concentration of As and Mn, a local $d$-weight maximum below $E_F$ accounts for only a few percent of the total one-particle DOS in this region.

The hole-doped nature of the valence-band states is also corroborated by a comparison of the experimental band structure between (Ga,Mn)As with $T_C$ = 62 K (Fig. 2f) and nonmagnetic n-type GaAs (n-GaAs) doped with Si at concentration of $10^{19}$ cm$^{-3}$ (Fig. 2h). The doped electron concentration of $10^{19}$ cm$^{-3}$ results in the surface depletion layer with a thickness of several tens of nm, which is much thicker than the probing depth with the He I$\alpha$ line in ARPES measurements. Hence, when discussing the binding energy in n-GaAs one needs to take into account the band bending effect at the surface, which is ineffective for metallic (Ga,Mn)As with a depletion layer less than 1 nm. By shifting the chemical potential of n-GaAs upward by ~0.9 eV, the band dispersion of n-GaAs fits well with that of (Ga,Mn)As (Fig. 2g). Taking into account that the sum of bandgap and SO splitting at the Γ point is ~1.85 eV, it is inferred that there is an upward surface band bending of ~0.95 eV (= 1.85-0.9 eV) in n-GaAs, in agreement with the reported Fermi-level pinning position of ~0.8 eV.[38] The effective mass of band B obtained by ARPES is 0.08 ± 0.02 $m_0$ ($m_0$: the free-electron mass), in good agreement with the calculated effective mass of the SO band (LH and HH bands should have a much larger effective mass)[39].

Figure 2i displays the contour plot of ARPES-spectral intensity at several binding energies from $E_F$ to 0.6 eV. We find that the intersection of band B has a circular shape indicative of the SO-band nature, supporting the above interpretation that the Fermi level of (Ga,Mn)As is deeply embedded in the valence band. On the other hand, we do not observe the HH and LH bands in Fig. 2, which is likely due to the intensity reduction by the disorder in sample as well as the photoemission matrix-element effect[23], as inferred from the strong orbital dependence of the calculated photoelectron intensity in GaAs[40] (see Supplementary Information). A part of the loss in the spectral intensity may be accounted for by the presence of a Coulomb gap discussed below.

**Coulomb gap.** Having established the position of the Fermi energy in the As 4$p$ valence band in (Ga,Mn)As, the next question is how the metallic state evolves from pristine GaAs by Mn doping. To address this question, we performed systematic ARPES measurements on a variety of samples with different $T_C$'s. Figure 3a shows a set of ARPES spectra at the $\Gamma$ point revealing a rounded shape in the vicinity of $E_F$ unlike typical metals such as gold (Au) which shows a sharp Fermi-edge cut-off as seen in Fig. 3a. This rounded feature of spectrum provides a new evidence for a depression of one-particle DOS at $E_F$ in (Ga,Mn)As, so far seen by tunnelling spectroscopy and assigned to the electron-electron correlation in the vicinity to a disorder-driven MIT[41-43].

According to the Altshuler-Aronov theory[27,44], the interaction-induced correction to single-particle DOS in a three-dimensional *metal* assumes the form $\delta\nu(\varepsilon) = a + b|\varepsilon|^{1/2}$ at $k_B T < |\varepsilon| < \hbar/\tau < E_F$, where $\varepsilon$ is the quasiparticle energy with respect to $E_F$, and $\tau$ is the momentum relaxation time. The sign and magnitude of $a$ and $b$ are determined by mutually compensating interference effects in the singlet and triplet particle-hole diffusion channels[45-47]. However, spin-disorder scattering in a paramagnetic phase or large band spin-splitting in a ferromagnetic phase destroy the interference in the triplet channel, which according to the perturbation[46,47] and dynamic renormalization group theory[45,48], should result in a sizable decrease of DOS at $E_F$, *i.e.*, $a < 0$, $b > 0$, and $\nu(0) \rightarrow 0$ on approaching the MIT from the metal side.

As shown in Fig. 3a, a numerically simulated ARPES spectrum with taking into account the expected form of the DOS depression, $\delta\nu(\varepsilon) = a + b|\varepsilon|^{1/2}$, where $a < 0$ and $b > 0$ reproduces satisfactorily the experimental spectrum up to ~0.1 eV relative to $E_F$ (Fig. 3b). This value coincides well with the energy at which the tunnelling DOS starts to be depleted[43] and also the optical conductivity is peaked[49], suggesting that the suppression of DOS near $E_F$ is an inherent feature of ferromagnetic (Ga,Mn)As. We observe a monotonic decrease of the near-$E_F$ spectral weight upon decreasing $T_C$. This is more clearly visible by plotting the DOS at $E_F$, *i.e.*, $\nu(0)$ as a function of $T_C$ (Fig. 3b), demonstrating a close relationship between the proximity to MIT, $T_C$, and $\nu(0)$. Accordingly, an effect from the soft Coulomb gap should be considered when discussing the one-particle excitations. The present ARPES results thus unambiguously demonstrate that (Ga,Mn)As with high $T_C$ possesses the metallic electronic structure with the Fermi level in the valence band, and is strongly influenced by the conjunction of correlation and disorder effects.

Magnetic linear dichroism. We now turn our attention to the "magnetic linear dichroism" (MLD) effect in the ARPES intensity of the valance band in (Ga,Mn)As. According to Figs. 4a and 4b the symmetry of the ARPES intensity with respect to the $\Gamma$ point shows a magnetization-direction dependence in (Ga,Mn)As with $T_C = 101$ K. As shown in Fig. 4a, the near-$E_F$ photoelectron intensity of the SO band in the ferromagnetic phase ($T = 30$ K) in the $k_{[\bar{1}\bar{1}0]}$ region (left-hand side, negative $k$ values) is stronger than that in the $k_{[110]}$ region (right-hand side) when we magnetize the sample along the [100] direction as illustrated in Fig. 4c (see also Figs. 2d and 2f). On the other hand, when we reverse the magnetization direction to [$\bar{1}$00] the photoelectron intensity in the positive $k$ region becomes more dominant (Fig. 4b). The intensity difference between two opposite magnetization cases is estimated to be 2-8% (depending on the binding energy) by taking account of background contribution of ~70 %. When we overlaid the band dispersion for two magnetization cases, we recognize a finite difference in their energy positions as shown in Fig. 4d. Interestingly, when we cool down the sample without external magnetic field, the intensity asymmetry becomes less clear (Fig. 4e). In addition, the asymmetry is not seen in nonmagnetic n-GaAs (Fig. 4f).

In order to determine the origin of this MLD we note that for the valence-band photoemission in question, photon-induced transitions occur between the extended valence band states (SO band in our case) and the free-electron-like states which propagate towards the detector in the vacuum while damped inside the crystal (the one-step model). Since there is no periodicity in the direction normal to the surface, $k_z$ is not conserved. Therefore valence states with various $k_z$ can contribute to the magnitude of emitted photocurrent at given $E_{kin}$ and $k_{//}$ values. Hence, following the method elaborated previously[50], we have calculated for a given orientation of $\boldsymbol{k}_{//}$ and energy $E_{SO}$ in the SO band the absorption coefficient for linearly polarized light considering electric-dipole transitions from the SO band split by the *p-d* exchange and spin-orbit interactions to spin-degenerate free-electron states. Since only $\boldsymbol{k}_{//}$ is conserved, the absorption coefficient for four possible transitions is a product of corresponding DOS $\nu(E_{SO}) = k_{//}^2 (2\pi^2 dE_{SO}/dk_{//})^{-1}$ at $k_z = 0$ and the square of the absolute value of the matrix element averaged over $k_z$ values. No MLD is found if the valence band is described within the six band Luttinger model[50]. In contrast, allowing for the presence in the 6×6 $kp$ Hamiltonian of terms brought about by the inversion asymmetry[51,52], MLD of characteristics observed experimentally shows up. There are three such contributions in the case under consideration: terms proportional to $k^3$, originally introduced by Dresselhaus for zinc-blende crystals, and two linear terms in $k$ components, appearing under biaxial and shear strain, respectively. In the absence of the exchange splitting, time reversal symmetry leads to the same magnitude of photoemission for **k** and -**k** if transitions from both spin subbands are taken into account. However, this is no longer the case if time reversal symmetry is broken by non-zero magnetization.

Figure 5 shows the computed magnitude of MLD = $(A_{k[110]} - A_{k[\bar{1}\bar{1}0]})/A_{k[110]}$ at **M** ∥ [100] for the SO band as well as for $kp$ parameters specified in Methods[11,15,18,51] and for the expected values of strain[2] $\varepsilon_{xx} = -0.4\%$ and $\varepsilon_{xy} = 0.1\%$ as well as for the magnitude exchange splitting of the valence band at $\Gamma_8$ point $\Delta_v = 180$ meV, corresponding to the saturation magnetization for $x = 0.05$. Since the character of light polarization is uncertain under our experimental conditions, the calculations have been performed for both s and p polarization. We have found that for the employed parameters, the $k^3$ contribution dominates. As seen, the computed magnitude of the average value of MLD over the two polarization is in accord with experimental results, *i.e.*, the theory predicts properly the sign, magnitude, and energy width of the effect as well as its symmetry, $A_{k[110]}(\mathbf{M} \parallel 100) = A_{k[\bar{1}\bar{1}0]}(\mathbf{M} \parallel \bar{1}00)$. It is worth noting that to resolve directly spin splitting and $k$ asymmetry of the valence-band subbands ultrahigh-resolution spin-resolved ARPES measurements would be necessary.

## Conclusions

To conclude, our ARPES studies of (Ga,Mn)As resolve particularly well the band split off by spin-orbit coupling (*i.e.*, $\Gamma_7$ valence band subband at $k = 0$). Since the magnitude of spin-orbit splitting is known quite precisely and varies little across the arsenide family of III-V compounds, we have been able to evaluate quite accurately the position of the Fermi level with respect to the top of the valence band. The obtained Fermi energy of about -0.3 eV in (Ga,Mn)As with $T_C$ of the order of 100 K is consistent with the *p-d* Zener model[16,50]. At the same time our data point to the presence of a substantial depression in DOS at $E_F$. We assign this observation, in accord with tunnelling studies[41-43], to electron correlation in disordered systems, which result in a depletion of DOS for the universality class in question[45-48]. We have also detected magnetic linear dichroism that results from inversion asymmetry of zinc-blende structure. Our computations within *kp* theory shows that a dominant contribution to the magnitude of MLD in the SO band comes from cubic $k^3$ terms. It would be interesting to check their contribution to spin-orbit torque analysed so-far considering only linear terms in *k* (refs. 13-15).

## METHODS

**Sample preparation and ARPES experiments.** 25-nm thick $Ga_{1-x}Mn_xAs$ films with *x* between 0.05 and 0.07 were grown at substrate temperature between 230 and 245°C on n$^+$-GaAs (001) substrate through a 100-nm thick n$^+$-GaAs buffer layer grown at ~560°C by molecular beam epitaxy (MBE). After the growth, we immediately transferred the films to a portable ultrahigh-vacuum (UHV) chamber equipped with non-evaporating getter and ion pumps, disconnected the portable chamber from the MBE system, and then connected to the ARPES system with keeping the UHV condition better than $5\times10^{-10}$ Torr. All the sample-transfer procedures were carried out within one hour. ARPES measurements were performed with the MBS-A1 electron analyzer equipped with a high-intensity He plasma discharge lamp. We use the He-Iα resonance line (photon energy: $hv = 21.218$ eV) to excite photoelectrons. The energy resolution for the ARPES measurements was set at 15-40 meV. The sample temperature was kept at 30 K during the measurements.

**Calculations.** A tight-binding method described previously[53] was employed to determine the band structure $E(\boldsymbol{k})$ of GaAs and $Ga_{0.95}Mn_{0.05}As$ in the whole Brillouin zone. Density of states and matrix elements for electric-dipole optical transitions between split-off band ($\Gamma_7$ at $k = 0$) and free electron states were computed from the six band Luttinger-Kohn $kp$ theory with the standard values[50] of the Luttinger parameters, the $p$-$d$ exchange integral β, and the elastic moduli $c_{ij}$, taking additionally into account odd in $k$ terms resulting from inversion asymmetry of the zinc-blende structure. The numerical values of $kp$ parameters ($b_{41}^{v7v7} = -58.71$ eVÅ$^2$, $b_{41}^{v8v8} = -81.93$ eVÅ$^2$, and $b_{41}^{v8v7} = -101.9$ eVÅ$^2$), describing the dominant $k^3$ terms, are taken from ref. 52. There are additional two contributions to the six band $kp$ Hamiltonian linear in $k$, contributing also to the sector of the $kp$ Hamiltonian describing the split-off band $H_{v7v7}$. These terms are generated by shear $\varepsilon_{ij}$ and diagonal $\varepsilon_{ii}$ strain tensor components that enters into $H_{v7v7}$ via **σ·φ** and **σ·ψ**, respectively[51], where $\sigma_i$ are Pauli matrices, $\varphi_x = k_y\varepsilon_{xy} - k_z\varepsilon_{xz}$, and $\psi_x = k_x(\varepsilon_{yy} - \varepsilon_{zz})$ (and cyclic permutations). The form of the six band Hamiltonian involving $\varphi_i$ and the corresponding deformation potential constant $C_4/\hbar = -2.2\times10^6$ m/s were given previously[18]. The magnitude of the deformation potential describing the term with $\psi_i$ is unknown. Following a previous approach[15], the same value of the deformation potential and also the same form of the Hamiltonian (with $\varphi_i$ replaced by $\psi_i$) are adopted. In the case under consideration $\varepsilon_{xy} \neq 0$, $\varepsilon_{yz} = \varepsilon_{zx} = 0$, and $\varepsilon_{xx} = \varepsilon_{yy} = -\varepsilon_{zz}c_{11}/2c_{12}$, where $z$ axis is taken along the growth direction[50,54].

**References**

1. Ohno, H. *et al*. (Ga,Mn)As: a new diluted magnetic semiconductor based on GaAs. *Appl. Phys. Lett.* **69**, 363-365 (1996).

2. Dietl, T. & Ohno, H. Dilute ferromagnetic semiconductors: Physics and spintronic structures. *Rev. Mod. Phys.* **86**, 187-251 (2014).

3. Jungwirth, T. *et al*. Spin-dependent phenomena and device concepts explored in (Ga,Mn)As. *Rev. Mod. Phys.* **86**, 855-896 (2014).

4. Jungwirth, T. *et al*. Character of states near the Fermi level in (Ga,Mn)As: Impurity to valence band crossover. *Phys. Rev. B* **76**, 125206 (2007).

5. Samarth, N. Battle of the bands. *Nat. Mater*. **11**, 360-361 (2012).

6. Ohno, H. *et al*. Electric-field control of ferromagnetism. *Nature* **408**, 944-946 (2000).

7. Boukari, H. *et al*. Light and electric field control of ferromagnetism in magnetic quantum structures. *Phys. Rev. Lett.* **88**, 207204 (2002).

8. Kohda, M., Ohno, Y. Takamura, K. Matsukura, F. & Ohno, H. A spin Esaki diode. *Jpn. J. Appl. Phys.* **40**, L1274-76 (2001).

9. Johnston-Halperin, E. *et al*. Spin polarized Zener tunneling in (Ga,Mn)As. *Phys. Rev. B* **65**, 041306 (2002).

10. Mattana, R. *et al*. Electrical detection of spin accumulation in a p-type GaAs quantum well. *Phys. Rev. Lett.* **90**, 166601 (2003).

11. Yamanouchi, M., Chiba, D., Matsukura, F. & Ohno, H. Current-induced domain-wall switching in a ferromagnetic semiconductor structure. *Nature* **428**, 539-542 (2004).

12. Chiba, D. *et al*. Magnetization vector manipulation by electric field. *Nature* **455**,

515-518 (2008).

13. Chernyshov, A. *et al*. Evidence for reversible control of magnetization in a ferromagnetic material by means of spin-orbit magnetic field. *Nat. Phys.* **5**, 656-659 (2009).

14. Endo, M., Matsukura, F. & Ohno, H. Current induced effective magnetic field and magnetization reversal in uniaxial anisotropy (Ga,Mn)As. *Appl. Phys. Lett.* **97**, 222501 (2010).

15. Kurebayashi, H. *et al. Nat. Nanotechnol.* **9**, 211 (2014).

16. Dietl, T., Ohno, H., Matsukura, F., Cibert, J. & Ferrand, D. Zener model description of ferromagnetism in zinc-blende magnetic semiconductors, *Science* **287**, 1019-1022 (2000).

17. Zemen, J., Kučera, J. Olejník, K & Jungwirth, T. Magnetocrystalline anisotropies in (Ga,Mn)As: Systematic theoretical study and comparison with experiment. *Phys. Rev. B* **80**, 155203 (2009).

18. Stefanowicz, W. *et al*. Magnetic anisotropy of epitaxial (Ga,Mn)As on (113)*A* GaAs. *Phys. Rev. B* **81**, 155203 (2010).

19. Bihler, C. *et al.* $Ga_{1-x}Mn_xAs$/piezoelectric actuator hybrids: A model system for magnetoelastic magnetization manipulation. *Phys. Rev. B* **78**, 045203 (2008).

20. Bernevig, B. A. & Vafek, O. Piezo-magnetoelectric effects in p-doped semiconductors. *Phys. Rev. B* **72**, 033203 (2005).

21. Gould, C. *et al.* Tunneling anisotropic magnetoresistance: A spin-valve like tunnel magnetoresistance using a single magnetic layer. *Phys. Rev. Lett.* **93**, 117203 (2004).

22. Edmonds, K., van der Laan, G. & Panaccione, G. Electronic structure of

(Ga,Mn)As seen by synchrotron radiation. *Semicond. Sci. Technol.* **30**, 043001 (2015).

23. Kobayashi, M. *et al.* Unveiling the impurity band induced ferromagnetism in the magnetic semiconductor (Ga,Mn)As. *Phys. Rev. B* **89**, 205204 (2014).

24. Di Marco, I. *et al*. Electron correlations in $Mn_xGa_{1-x}As$ as seen by resonant electron spectroscopy and dynamical mean field theory. *Nat. Commun*. 4:2645 doi: 10.1038/ncomms3645 (2013).

25. Gray, A. X. *et al.* Bulk electronic structure of the dilute magnetic semiconductor $Ga_{1-x}Mn_xAs$ through hard X-ray angle-resolved photoemission. *Nature Mater.* **11**, 957-962 (2012).

26. Okabayashi, J. *et al.* Angle-resolved photoemission study of $Ga_{1-x}Mn_xAs$. *Phys. Rev. B* **64**, 125304 (2001).

27. Al'tshuler, B. L. & Aronov, A. G. Contribution to the theory of disordered metals in strongly doped semiconductors, *Zh. Eksp. Teor. Fiz.* **77**, 2028-2044 (1979) [*Sov. Phys. JETP* **50**, 968-976 (1979)].

28. Edmonds, K. W. *et al*. Magnetic linear dichroism in the angular dependence of core-level photoemission from (Ga,Mn)As using hard X rays, *Phys. Rev. Lett.* **107**, 197601 (2011).

29. van der Laan, G. Magnitude of core-hole orbital moments from magnetic linear dichroism in the angular dependence of photoemission, *Phys. Rev. B* **55**, 3656-3662 (1997).

30. Ohno, H. Making nonmagnetic semiconductors ferromagnetic. *Science* **281**, 951-956 (1998).

31. Shimizu, H., Hayashi, T., Nishinaga, T. & Tanaka, M. Magnetic and transport

properties of III-V based magnetic semiconductor (Ga,Mn)As: Growth condition dependence. *Appl. Phys. Lett.* **74**, 398-400 (1999).

32. Yu, K. M. *et al*. Effect of the location of Mn sites in ferromagnetic $Ga_{1-x}Mn_xAs$ on its Curie temperature. *Phys. Rev. B* **65**, 201303(R) (2002).

33. Blinowski, J. & Kacman, P. Spin interaction of interstitial Mn ions in ferromagnetic GaMnAs. *Phys. Rev. B* **67**, 121204(R) (2003).

34. Jungwirth, T. *et al*. Prospects for high temperature ferromagnetism in (Ga,Mn)As semiconductors. *Phys. Rev. B* **72**, 165204 (2005).

35. Edmonds, K. W. *et al*. Mn interstitial diffusion in (Ga,Mn)As. *Phys. Rev. Lett.* **92**, 037201 (2004).

36. Chiang, T.-C. *et al*. Angle-resolved photoemission studies of GaAs (100) surface grown by molecular-beam epitaxy. *Phys. Rev. B* **27**, 4770-4778 (1983).

37. Kumigashira, H. *et al*. High-resolution angle-resolved photoemission study of LaSb. *Phys. Rev. B* **58**, 7675-7680 (2010).

38. Adachi, S. GaAs and related materials: Bulk semiconducting and superlattice properties (World Scientific, 1994).

39. Lowney, J. R. & Kahn, A. H. Valence-band effective masses of GaAs. *J. Appl. Phys.* **64**, 447-450 (1988).

40. Larsen, P. K., Neave, J. H. & Joyce, B. A. Angle-resolved photoemission from As-stable GaAs (001) surfaces prepared by MBE. *J. Phys. C: Solid State Phys*. **14**, 167-192 (1981).

41. Chun, S. H., Potashnik, S. J., Ku, K. C., Schiffer, P. & Samarth, N. Spin-polarized tunnelling in hybrid metal-semiconductor magnetic tunnel junctions. Phys. Rev. B **66**, 100408 (2002).

42. Pappert, K. *et al*. Magnetization-switched metal-insulator transition in (Ga,Mn)As tunnel device. *Phys. Rev. Lett.* **97**, 186402 (2006).

43. Richardella, A. *et al.* Visualizing critical correlations near the metal-insulator transition in $Ga_{1-x}Mn_xAs$. *Science* **327**, 665-669 (2010).

44. Kobayashi, M., Tanaka, K., Fujimori, A., Ray, Sugata & Sarma, D. D. Critical test for Altshuler-Aronov theory: Evolution of the density of states singularity in double perovskite $Sr_2FeMoO_6$ with controlled disorder. *Phys. Rev. Lett.* **98**, 246401 (2007).

45. Finkelstein, A. M. Electron liquid in disordered conductors. *Soviet Sci. Rev. A Phys.* **14**, 1-101 (1990).

46. Altshuler, B. L. & Aronov, A. G. in *Electron-Electron Interactions in Disordered Systems*, edited by A. L. Efros, and M. Pollak (North-Holland, Amsterdam, 1985), p. 41.

47. Lee, P. A. & Ramakrishnan, T. V. Disordered electronic systems. *Rev. Mod. Phys.* **57**, 287-337 (1985).

48. Wojtowicz, T., Dietl, T., Sawicki, M., Plesiewicz, W. & Jaroszyński, J. Metal-insulator transition in semimagnetic semiconductors, *Phys. Rev. Lett.* **56**, 2419-2422 (1986).

49. Burch, K. S. *et al*. Impurity band conduction in a high temperature ferromagnetic semiconductor. *Phys. Rev. Lett.* **97**, 087208 (2006).

50. Dietl, T., Ohno, H. & Matsukura, F. Hole-mediated ferromagnetism in tetrahedrally coordinated semiconductors. *Phys. Rev. B* **63**, 195205 (2001).

51. Ivchenko, E. L. & Pikus, G. *Superlattices and Other Heterostructures*. Springer Series in Solid-State Sciences, vol. 110 (Springer, Berlin, 1995).

52. Winkler, R. *Spin–orbit Coupling Effects in Two-Dimensional Electron and Hole Systems.* Springer Tracts in Modern Physics, vol. 191 (Springer, Berlin, Heidelberg, 2003).

53. Oszwałdowski, R. Majewski, J. M. & Dietl, T. Influence of band structure effects on domain-wall resistance in diluted ferromagnetic semiconductors. *Phys. Rev. B* **74**, 153310 (2006).

54. Birowska, M., Śliwa, C., Majewski, J. A. & Dietl, T. Origin of bulk uniaxial anisotropy in zinc-blende dilute magnetic semiconductors. *Phys. Rev. Lett.* **108**, 237203 (2012).

55. Sankowski, P. *et al.*, Spin-dependent tunnelling in modulated structures of (Ga,Mn)As. *Phys. Rev. B* **75**, 045306 (2007).

**Acknowledgements**

We thank K. Honma for his assistance in ARPES measurements and S. Miyakozawa for his assistance in magnetization measurements. This work was supported by Japan Society for the Promotion of Science (KAKENHI 23224010, 26287071), Ministry of Education of Culture, Sports, Science and Technology in Japan (KAKENHI 26103002), European Research Council (#227690), National Center of Science in Poland (2011/02/A/ST3/00125), and World Premier International Research Center, Advanced Institute for Materials Research.

## Author Contributions

The work was planned and proceeded by discussion among S.S., F.M., T.D, H.O., and T.T., L.C., F.M., and H.O. carried out the samples' growth and their characterization. S.S., T.S., and T.T. performed ARPES measurements. T.D. contributed to the interpretation and analysis of experimental data. R.O performed tight-binding computations. S.S., F.M., and T.D. finalized the manuscript with inputs from all the authors.

## Additional Information

**Supplementary information** accompanies this paper at http://…..

**Competing financial interests:** The authors declare no competing financial interests.

**FIGURE LEGENDS**

**Figure 1 | Valence-band structure of heavily Mn-doped GaAs. a,** Magnetization $M$ of (Ga,Mn)As with $x = 0.05$ grown at 245°C measured under magnetic field $\mu_0 H$ of 0.8 mT along the in-plane $\langle 100 \rangle$ orientation. **b,** Bulk and surface Brillouin zones of (Ga,Mn)As, together with the emission plane in ARPES measurements along the $\overline{\Gamma}\overline{X}$ cut (green shade). **c,** Experimental band structure along the $\overline{\Gamma}\overline{X}$ cut for (Ga,Mn)As ($T_C$ = 101 K) at $T$ = 30 K measured with the He-Iα resonance line, obtained by plotting the ARPES intensity in the linear scale as a function of binding energy and wave vector. Black dashed curves are a guide for eyes to trace the band dispersions. Inset shows the same plot with an enhanced colour contrast in the area enclosed by red rectangle. **d,** Experimental band structure obtained by plotting the second derivative intensity of MDCs. Bands are labeled with A-D. **e,** Comparison of experimental band dispersions of (Ga,Mn)As with the tight-binding calculation for pristine GaAs at $k_z = 0$ (ΓKX plane; solid curves) and $k_z = 2\pi/a_0$ (XKΓ plane; dashed curves). HH, LH, and SO denote the heavy-hole, light-hole, and split-off bands, respectively. The experimental band dispersions are extracted by tracing the peak position in the second-derivative intensity of MDCs (open circles) and ARPES spectra (filled circles).

**Figure 2 | Evidence for the metallic valence band in (Ga,Mn)As. a,** Plot of near-$E_F$ ARPES intensity around the Γ point for (Ga,Mn)As with $T_C$ = 101 K as a function of binding energy and wave vector $k$. **b,** Raw ARPES spectrum (red curve) measured at a cut shown by solid green line in **a** and the corresponding background spectrum (black curve) obtained by integrating ARPES-spectral intensity over a wide $k$ region (0.4 Å$^{-1}$) around the Γ point. Background-subtracted ARPES spectrum is shown

with blue curve. **c**, **d**, A set of background-subtracted ARPES spectra and corresponding intensity plot, respectively, for (Ga,Mn)As with $T_C$ = 101 K. White curve is MDC at $E_F$. **e**, **f**, Same as **c** and **d** but for (Ga,Mn)As with $T_C$ = 62 K. **g**, Comparison of the band dispersions among (Ga,Mn)As with $T_C$ = 101 K and 62 K and n-GaAs, extracted from the peak position of MDCs, together with the band-structure calculation (same as Fig. 1e). Representative fitting result to the MDCs (red curve) with two Lorentzians (black curve) is shown in the bottom. **h**, ARPES intensity around the Γ point of n-GaAs. **i**, Intensity contour plots for (Ga,Mn)As with $T_C$ = 62 K as a function of in-plane wave vector at several energy slices from $E_F$ to 0.6 eV.

**Figure 3 | Evolution of metallic states and soft Coulomb gap. a**, Comparison of raw ARPES spectra around the Γ point (integrated over ±0.08 Å$^{-1}$ centered at the Γ point) for (Ga,Mn)As with various $T_C$'s (35, 60, 62, 101, and 112 K) and nonmagnetic n-GaAs. The intensity is normalized to the spectral weight integrated over 0-1 eV. Numerical simulation of the ARPES spectrum (blue solid curve) for (Ga,Mn)As with $T_C$ = 101 K by taking the spectral DOS with a soft Coulomb gap (dotted curve). The simulation was performed with a linearly decreasing DOS with square root energy dependence of the gap[27], multiplied by the Fermi-Dirac distribution function at $T$ = 30 K and convoluted with the resolution function (energy resolution 15 meV). Shaded area highlights the energy region of the gap. ARPES spectrum of gold (Au) is also shown for comparison. **b**, Spectral DOS at $E_F$ plotted as a function of $T_C$. The DOS is obtained by integrating the ARPES-spectral intensity within ±40 meV of $E_F$. Insets show schematics of the DOS in (Ga,Mn)As with high and low $T_C$.

**Figure 4 | Magnetization dependence of ARPES intensity in (Ga,Mn)As. a, b,** Near-$E_F$ ARPES intensity at $T = 30$ K of (Ga,Mn)As with $T_C = 101$ K, magnetized along the [100] and [$\bar{1}$00] directions, respectively. Calculated band dispersion of (Ga,Mn)As with incorporating the exchange splitting[55] is shown with yellow curves for comparison. **c,** Experimental geometry of sample axes, magnetization direction, and incident photons. Solid and dashed rectangles indicate the plane of incidence (110) and the emission plane of photoelectrons (1$\bar{1}$0), respectively. **d,** Location of the energy band extracted from the peak positions of the momentum distribution curves (MDCs) in **a** and **b**. **e,** Same as **a** and **b** but with zero-field (z.f.) cooling. **f,** Near-$E_F$ ARPES intensity at $T = 30$ K of GaAs (n-type; Si-doped).

**Figure 5 | Theoretical magnitudes of magnetic linear dichroism generated by inversion asymmetry in ferromagnetic (Ga,Mn)As epilayers.** Here, MLD = ($A_{k[110]}$ - $A_{k[\bar{1}\bar{1}0]}$) /$A_{k[110]}$, where $A$ is a product of the square of the absolute value of the matrix element for electric dipole transitions average over $|k_z| \leq 0.2\pi/a_0$ and density of states at $k \parallel [110]$ and [$\bar{1}\bar{1}0$], respectively corresponding to a given energy in the split-off band at $k_z = 0$. The magnitude of the exchange splitting of the valence band $\Delta_v$ = 180 meV (at $\Gamma_8$ point) corresponds to the value of saturation magnetization $M$ for Mn content $x = 0.05$. Empty symbols are for $M \parallel$ [100] and for s and p light polarization (see Fig. 4**c**). The full symbols depict average values weighted by relative magnitudes of $A$ for s and p polarization. The sign of MLD is reversed for $M \parallel [\bar{1}00]$.

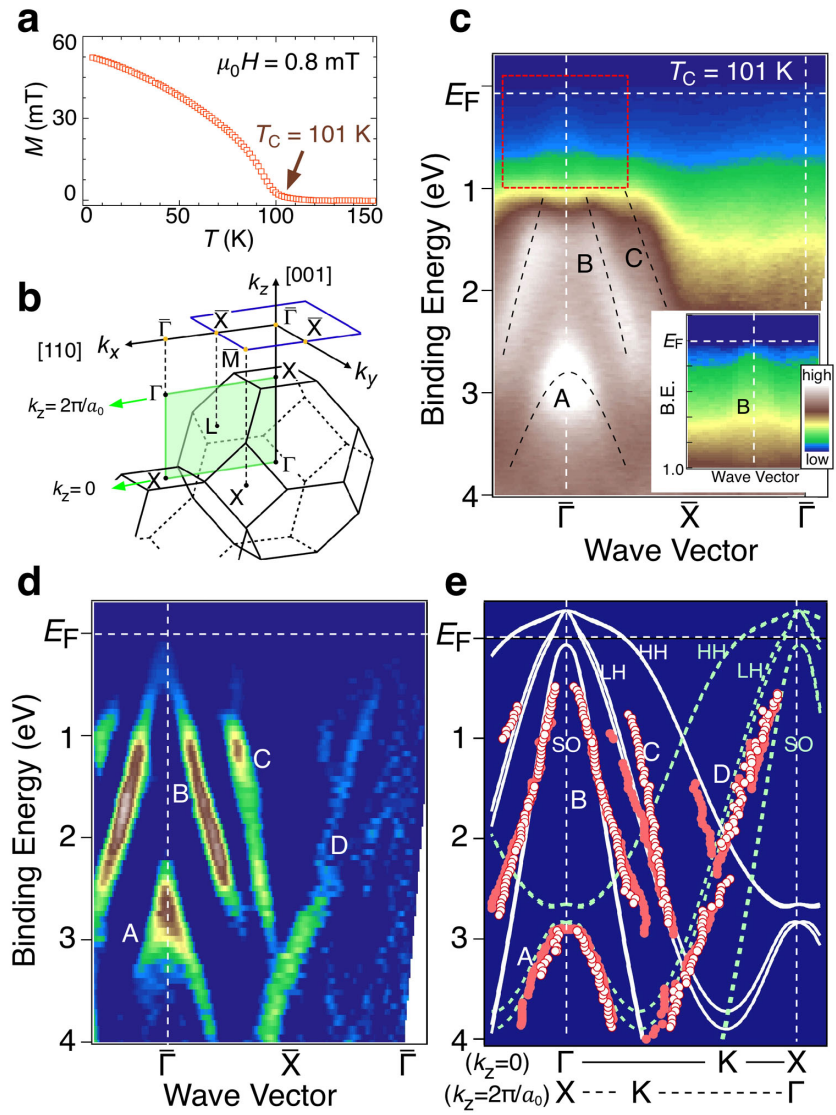

Figure 1

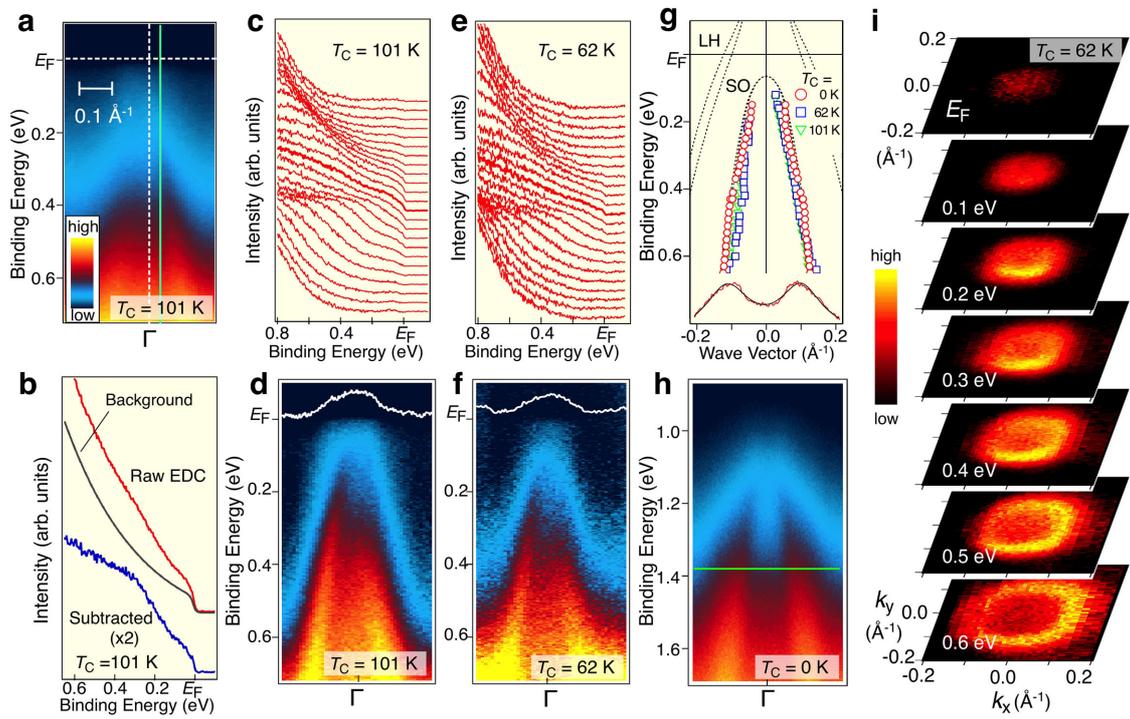

Figure 2

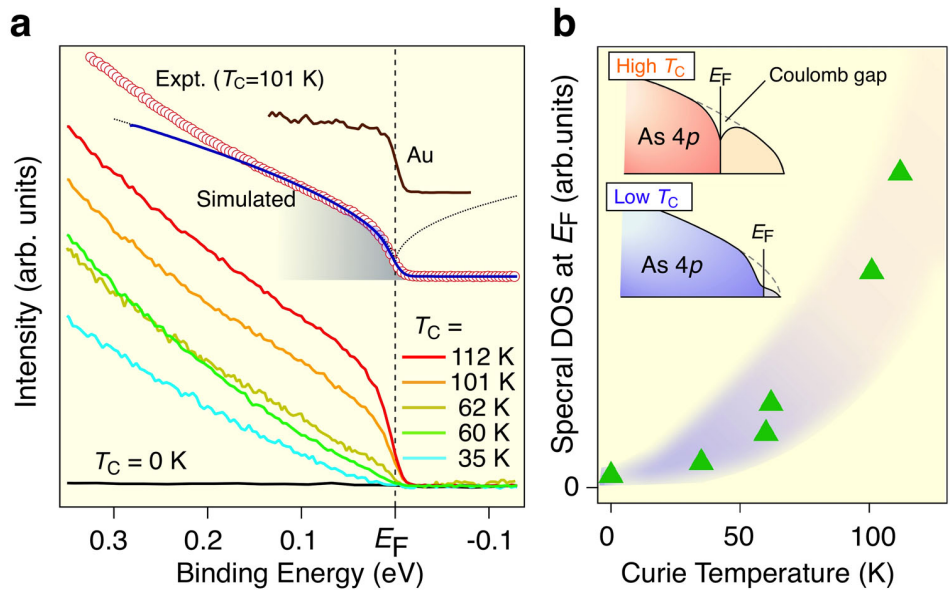

Figure 3

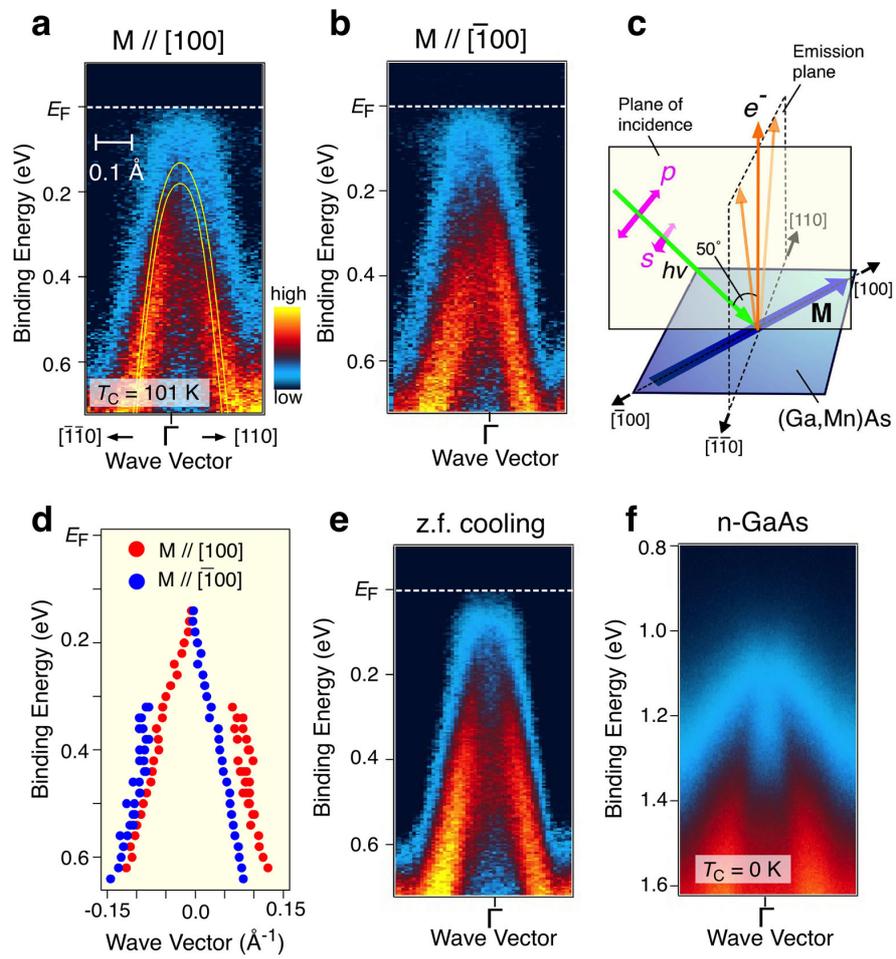

Figure 4

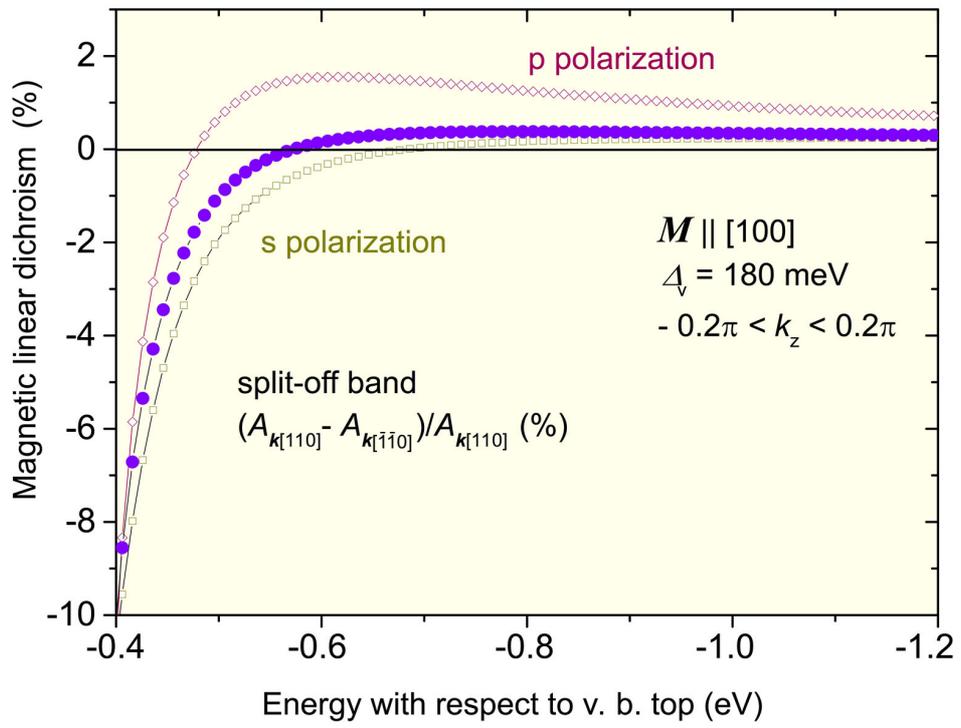

Figure 5

SUPPLEMENTARY INFORMATION

# Fermi level position, Coulomb gap, and Dresselhaus splitting in (Ga,Mn)As

S. Souma[1], L. Chen[1], R. Oszwałdowski[2], T. Sato[3], F. Matsukura[1,4,5], T. Dietl[1,6,7], H. Ohno[1,4,5], & T. Takahashi[1,3].

[1]*WPI Research Center, Advanced Institute for Materials Research, Tohoku University, 2-1-1 Katahira, Aoba-ku, Sendai 980-8577, Japan*

[2]*Department of Physics, South Dakota School of Mines and Technology, Rapid City, SD 57701, USA*

[3]*Department of Physics, 6-3 Aramaki Aza-Aoba, Aoba-ku, Tohoku University, Sendai 980-8578, Japan*

[4]*Center for Spintronics Integrated Systems, Tohoku University, 2-1-1 Katahira, Aoba-ku, Sendai 980-8577, Japan*

[5]*Laboratory for Nanoelectronics and Spintronics, Research Institute of Electrical Communication, Tohoku University, 2-1-1 Katahira, Aoba-ku, Sendai 980-8577, Japan*

[6]*Institute of Physics, Polish Academy of Sciences, aleja Lotników 32/46, PL-02-668 Warszawa, Poland*

[7]*Institute of Theoretical Physics, Faculty of Physics, University of Warsaw, ulica Pasteura 5, PL-02-093 Warszawa, Poland.*

1. Band assignment of GaAs and photoelectron matrix-element effect

We observed a strong selection rule of photoelectron intensity[40] (*i.e.* photoemission matrix-element effect) in the ARPES data of GaAs and (Ga,Mn)As. As displayed in Fig. S1**a**, when the polarization vector of monochromatized He-Iα resonance line points perpendicular to the measured $\overline{\Gamma}\overline{X}$ cut (see inset), both the light hole (LH) and heavy hole (HH) bands are well resolved, particularly in high binding-energy region, while the intensity of the split-off (SO) band appears to be very weak. On the other hand, when the polarization vector is aligned parallel to the $\overline{\Gamma}\overline{X}$ cut as shown in Fig. S1**b**, the SO band appear to be more visible, while the LH/HH bands become dimmer. This characteristic polarization dependence of ARPES intensity is consistent with the previous report of soft-x-ray ARPES on (Ga,Mn)As[23].

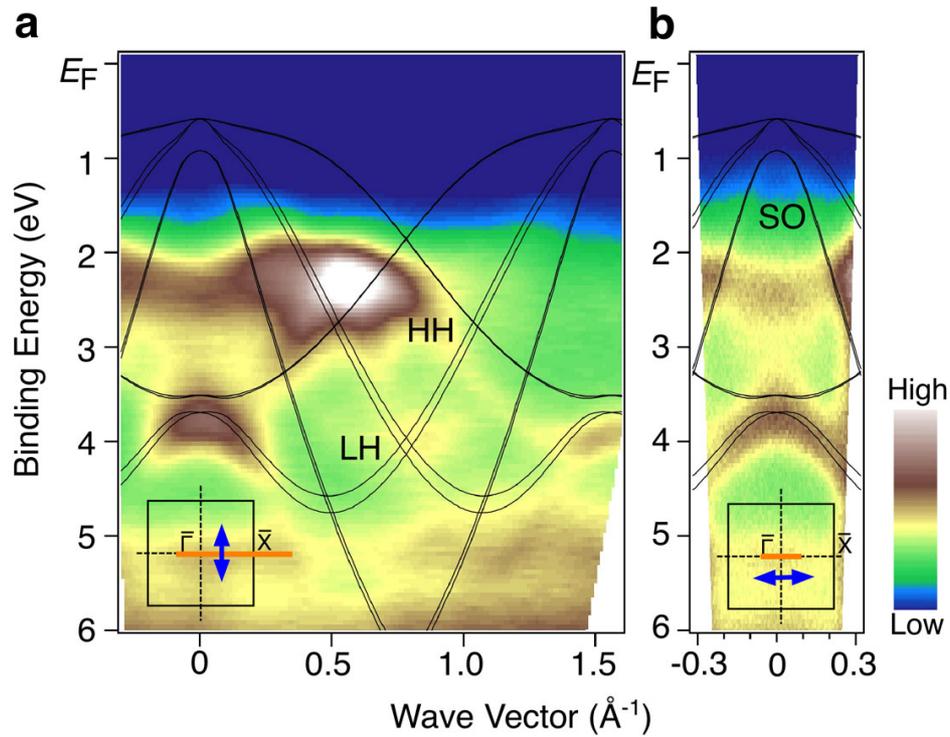

**Figure S1 | Light-polarization dependence of ARPES intensity of GaAs. a**, ARPES intensity of GaAs (n-type; Si-doped) measured along the ΓX cut by setting the polarization vector of incident light perpendicular to the measured cut (see arrow in inset). **b**, ARPES intensity of GaAs measured with the polarization vector parallel to the measured cut (inset). Calculated band structure of GaAs within the tight-binding approximation (same as Fig. 1e) is overlaid by solid curves in **a** and **b**.